\documentclass[aps,prl,twocolumn,superscriptaddress]{revtex4}

\usepackage[latin9]{inputenc}
\usepackage{amsmath}
\usepackage{amssymb}
\usepackage{graphicx}
\usepackage{makecell}
\usepackage{array}
\usepackage{CJK}

\usepackage[unicode=true,pdfusetitle,
 bookmarks=false,
 breaklinks=false,pdfborder={0 0 1},backref=false,colorlinks=false]
 {hyperref}
\hypersetup{
 bookmarksnumbered=false,bookmarksopen=false}

\makeatletter
\@ifundefined{textcolor}{}
{%
 \definecolor{BLACK}{gray}{0}
 \definecolor{WHITE}{gray}{1}
 \definecolor{RED}{rgb}{1,0,0}
 \definecolor{GREEN}{rgb}{0,1,0}
 \definecolor{BLUE}{rgb}{0,0,1}
 \definecolor{CYAN}{cmyk}{1,0,0,0}
 \definecolor{MAGENTA}{cmyk}{0,1,0,0}
 \definecolor{YELLOW}{cmyk}{0,0,1,0}
}

\usepackage{color}

\newcommand{\ehbar}{\hbar_{\mathrm{eff}}}

\@ifundefined{textcolor}{}{%
 \definecolor{BLACK}{gray}{0}
 \definecolor{WHITE}{gray}{1}
 \definecolor{RED}{rgb}{1,0,0}
 \definecolor{GREEN}{rgb}{0,1,0}
 \definecolor{BLUE}{rgb}{0,0,1}
 \definecolor{CYAN}{cmyk}{1,0,0,0}
 \definecolor{MAGENTA}{cmyk}{0,1,0,0}
 \definecolor{YELLOW}{cmyk}{0,0,1,0}
}

\usepackage{txfonts}

\makeatother

\begin{document}

\title{Self-interaction induced phase modulation for directed current, energy diffusion and quantum scrambling in a Floquet ratchet system}

\author{Jiejin Shi}
\affiliation{School of Science, Jiangxi University of Science and Technology, Ganzhou 341000, China}

\author{Lihao Hua}
\affiliation{School of Science, Jiangxi University of Science and Technology, Ganzhou 341000, China}

\author{Wenxuan Song}
\email[]{swx240215@163.com}
\affiliation{School of Science, Jiangxi University of Science and Technology, Ganzhou 341000, China}

\author{Wen-Lei Zhao}
\email[]{wlzhao@jxust.edu.cn}
\affiliation{School of Science, Jiangxi University of Science and Technology, Ganzhou 341000, China}

\begin{abstract}

We investigate the wavepacket dynamics in an interacting Floquet system described by the Gross-Pitaevskii equation with a ratchet potential. Under quantum resonance conditions, we thoroughly examine the exotic dynamics of directed current, mean energy, and quantum scrambling, based on the exact expression of a time-evolving wavepacket. The directed current is controlled by the phase of the ratchet potential and remains independent of the self-interaction strength. Interestingly, the phase modulation induced by self-interaction dominates the quadratic growth of both mean energy and Out-of-Time-Ordered Correlators (OTOCs). In the quantum nonresonance condition, the disorder in momentum space, induced by the pseudorandom feature of the free evolution operator, suppresses the directed current at all times. Meanwhile, the disorder also leads to the dynamical localization of the mean energy and the freezing of quantum scrambling for initially finite time interval. The dynamical localization can be effectively manipulated by the phase, with underlying physics rooted in the different quasi-eigenenergy spectrum modulated by ratchet potential. Both the mean energy and OTOCs exponentially increase after long time evolution, which is governed by the classically chaotic dynamics dependent on the self-interaction. Possible applications of our findings on quantum control are discussed.
\end{abstract}
\date{\today}

\maketitle

\section{Introduction}
Physics induced by self-interaction, which describes the Kerr effect in nonlinear optics or the mean-field approximation of a Bose-Einstein condensate (BEC), has received considerable interest in the study of Kardar-Parisi-Zhang physics~\cite{SMU24prl}, quantum chaos~\cite{Graham05jpa,Guarneri17pre,WLZhao21prb,wlzhao16pra,wlzhao19jpa}, and quantum thermalization~\cite{SMU22epl,Haldar23epl,Banerjee24arx}. It has been found that nonlinear coupling can be used to control negative-positive mass transition of spin soliton in two-component BEC system~\cite{Meng22pra}. The self-interaction even leads to a topological phase transition, evidenced by the emergence of nonlinear Dirac cones~\cite{Bomantara17prb} and topological solitons~\cite{Tuloup20prb}, unveiling rich physics absent in non-interacting systems. Interestingly, time modulation of self-interaction has been employed to engineer wavepacket dynamics. For example, delta-kicking modulation of self-interaction can suppress the dispersive spreading of wavepackets~\cite{Goussev18pra}, results in the super-exponential increase of out-of-time-ordered correlators (OTOCs)\cite{WLZhao21prb,WLZhao23arx}, and the sub-diffusion of energy\cite{Duval22pra}. In recent years, the time-periodic modulation of self-interaction has served as a powerful tool to explore intriguing phenomena across different fields of physics.

The phase incorporated in external driven potentials has been exploited as an effective knob to manipulate the symmetry of Floquet systems, thereby offering potential applications in engineering the wavepacket's dynamics~\cite{JBGong02prl,JBGong01prl,Bitter17prl,Sentef20prr,Bai20pra,Downing23}, such as directed transport~\cite{Kenfack08prl,Hainaut18pra,ZQLi23pra,Poletti07pra} and information scrambling~\cite{FXChen24arx,Meier19pra,JHWang22prr,JHarris22prl}.
For example, the phase in ratchet potential governs both the direction and the magnitude of the directed current~\cite{Zhang15SB}. In addition, quasiperiodical modulation for the phase in kicking potential can be used to create the high-dimensional synthetic space~\cite{Casati89prl,Shepelyansky87,Tian11prl}, wherein fruitful physics, such as the Anderson metal-insulator transition~\cite{Lopez12prl,Cherroret14prl,Lopez13,Manai15prl} and the time-driven quantum phase~\cite{Hainaut18prl}, are experimentally achieved by using a variant of the kicked rotor model. The quasiperiodical modulation of the phase even facilitates the quantized Hall-like conductance of energy diffusion, rooted in quantum chaos, in kicked spin-$1/2$ rotor model~\cite{Chen14prl}. Notably, the adiabatic modulating on the phase in delta-kicking ratchet potential induces rich Floquet band topology, signatured by topological transport in momentum space~\cite{Derek12prl}.

In this context, we investigate both analytically and numerically the combined effects of self-interaction and phase modulation on directed transport, energy diffusion, and quantum scrambling, using a quantum kicked rotor model with a ratchet potential. Under quantum resonance condition, the mean momentum increases linearly with time, signaling directed current, with a growth rate independent of the self-interaction strength and governed by the ratchet potential phase. Both the mean energy and OTOCs increase quadratically with time, with their growth rates containing a term that combines the self-interaction strength and the sine function of the phase, indicating a phase modulation induced by nonlinearity. In the quantum nonresonance case, directed current is suppressed by disorder in momentum space. Both the mean energy and OTOCs exhibit a transition from dynamical localization to exponential growth as time evolves. The exponential growth of mean energy follows classically chaotic diffusion, indicating a quantum-to-classical transition. Interestingly, the dynamical localization can be effectively adjusted by the phase of the ratchet potential, as reflected by periodic oscillations or saturation in the mean energy for different phases. We uncover the underlying mechanism through the investigation of the quasienergy spectrum. Our findings have potential applications for engineering directed current, energy diffusion, and quantum scrambling by tuning the phase of the external field in interacting Floquet systems.

The paper is organized as follows. In Sec.~\ref{Sec-MResl} we describe the system. In Sec.~\ref{QuanReson}
, we show the directed current, energy diffusion and quantum scrambling in quantum resonance case. In Sec.~\ref{QuanNReson}, we discuss the wavepacket's dynamics in quantum nonresonance case. A summary is presented in Sec.~\ref{Sum}.

\section{Nonlinear QKR Model}\label{Sec-MResl}
The dimensionless Hamiltonian of the QKR model with self-interaction reads
\begin{equation}
\begin{aligned}\label{hmd}
{\rm H}=\frac{p^2}{2}+V_K(\theta)\sum\limits_n\delta(t-t_n)+\text{H}_{\text{I}}\sum\limits_n\delta(t-t_n) \:,
\end{aligned}
\end{equation}
\begin{align}
\begin{aligned}\label{vk}
V_K(\theta)=K\left[\sin(\theta)+\alpha\sin(2\theta+\phi)\right]\:,
\end{aligned}
\end{align}
and
\begin{align}
\begin{aligned}\label{Hi}
\text{H}_{\text{I}}=g|\psi(\theta)|^2\:,
\end{aligned}
\end{align}
where $p = -i\ehbar\partial/\partial\theta$ is the angular momentum operator, $\theta$ is the angle coordinate, satisfying communicate relation $[\theta,p]=i\ehbar$ with $\ehbar$ the effective Planck constant. The parameter $K$ denotes the strength of the kicking potential, $\phi$ and $\alpha$ determine the asymmetry and strength of the ratchet potential, respectively, $g$ is the nonlinearity interaction strength. The time $t_n$ is integer, i.e., $t_n = 1, 2 . . .$, indicating the kicking number. The eigenequation of angular momentum operator is $p|\varphi_n\rangle = p_n |\varphi_n \rangle$ with eigenvalue $p_n = n\ehbar$ and eigenstate $\langle \theta|\varphi_n\rangle=e^{in\theta}/\sqrt{2\pi}$. An arbitrary quantum state can be expanded with the complete  basis as $|\psi \rangle=\sum_n \psi_n |\varphi_n\rangle$. The Floquet operator consists of two components, namely $U=U_fU_K$, where the $U_f =\exp\left(-ip^2/2\ehbar\right)$ is the free evolution operator and the $U_K =\exp\left\{-i	\left[V_K(\theta)+g|\psi(\theta)|^2\right]/\ehbar\right\}$ is the kicking evolution operator. The evolution of a quantum state in one period from $t_n$ to $t_{n+1}$ depends on $|\psi(t_{n+1})\rangle = U|\psi(t_{n})\rangle $.

Our consideration of the ratchet potential is inspired by the fact that systems displaying rich physics can be realized in optical setups~\cite{Zhang15SB}. Here, we propose an optical experiment to simulate the quantum state evolution governed by the Hamiltonian in Eq.~\eqref{hmd}. It is well known that, under the paraxial approximation, the propagation of light is described by an equation mathematically equivalent to the Schr{\" o}dinger equation~\cite{Sharabi18,Prange89}, with the longitudinal dimension of light mimicking the time variable. Optical realization of kicked rotor model is implemented by a periodic sequence of multilayers of phase gratings~\cite{Fischer00pre,Rosen00JOS}, which has the advantage of the controllability of the kicking times by engineering the numbers of the layers of phase grating. Note that, to realize the delta kicks in time, both the sizes of the phase grating and Kerr media in the propagation direction should be much smaller than the period of the optical sequence. The propagation equation of light in such an optical system is discribed by the Hamiltonian in Eq.~\eqref{hmd}. Interestingly, the phase of the loss gratings can be precisely adjusted by etching the surface to different depths~\cite{Zhang15SB}, ensuring the achievement of ratchet potentials. The Kerr effect of media causes a intensity-dependent nonlinear term in Eq.~\eqref{hmd}. The mean value of the observables can be measured in the frequency domain of optics, enabling the observation of our findings.
Therefore, our finding is within reach of current experiments and may shed new light on the fundamental problems of quantum diffusion.

\section{Quantum resonance case} \label{QuanReson}

In the quantum resonance condition (i.e., $\ehbar=4\pi$), each element of the free evolution operator in momentum space is unity, i.e., $U_f(p_n)=\exp\left(-i{n}^2 2\pi\right)=1$. Consequently, it has no effect on the time evolution of the quantum state, leading to $U = U_K$. After arbitrary kicking period (i.e., $t=t_n$), the exact express of the quantum state takes the form $\psi(\theta,t_n)= U_K^{t_n}\psi(\theta,t_0)=\exp\left\{-i[V_K(\theta)+g|\psi(\theta,t_0)|^2]t_n/\ehbar \right\}\psi(\theta,t_0)$. The quantum resonance condition induces rich physics in various kicked rotor models. In the spinor kicked rotor, where ground hyperfine levels mimic pseudospin degrees of freedom, the system exhibits quantum walks in momentum space under resonance~\cite{Dadras19,Dadras18PRL,Summy16PRA}. In the double-kicked rotor, quantum resonance creates exotic quasienergy spectra, including flat bands~\cite{HLWang13PRE}, Dirac cones~\cite{Bomantara16PRE,Zhou14epjb}, and Hofstadter's butterfly~\cite{JWang08PRA}, laying the groundwork for topologically protected transport in generalized kicked rotor models~\cite{Zhoulw14,LWZhou23arx}. For the QKR with asymmetric potential, the ratchet effect leads to the directed current in momentum space~\cite{Lundh05PRL,Zhang15SB}, which opens the opportunity for controlling the wavepacket's dynamics~\cite{Kenfack08prl}.

The directed current in the ratchet effect arises from the different symmetries between the kicking potential and the initial state~\cite{Zhaowl14PRE,Lundh06PRE}. Without loss of generality, we choose an initial state with even symmetry, i.e., $\psi(\theta,t_0) = \cos(\theta)/\sqrt{\pi}$. Then, the wavefunction $\psi(\theta,t)$ in coordinate space is given by
\begin{equation}\label{QStateTn}
\psi(\theta,t)=
\exp\left\{-\frac{it}{4\pi}\left[V_K(\theta)+g\frac{\cos^2(\theta)}{\pi}\right]\right\}\frac{\cos(\theta)}{\sqrt{\pi}}\:.
\end{equation}
Based on this state, we can investigate analytically the time dependence of the momentum current $\langle p(t)\rangle=\sum_n p_n|\psi_n(t)|^2$, mean energy $\langle p^2(t)\rangle=\sum_n p_n^2|\psi_n(t)|^2$, and quantum scrambling $C(t)=-\langle [A(t),B]^2\rangle$~\cite{ZQi23,XDHu23,JWang21pre,Gribben24arx,Sharma24arx,Rozenbaum17prl}. Note that the OTOCs are defined by the average of the squared commutator, namely $C(t)=-\langle[A(t),B]^2\rangle$, where the operators $A(t)=U^{\dagger}(t)A U(t)$ and $B$ are evaluated in Heisenberg picture~\cite{Garc24prd}. The expectation value $\langle \cdot \rangle = \langle \psi(t_0) | \cdot | \psi(t_0) \rangle$ is taken with respect to the initial state~\cite{Li2017,Hashimoto2017,Mata2018,Zonnios22,WLZhao21prb,Pappalardi22SP,Khalouf24pra,SYWang23arx}.
We consider the case where $A=\exp\left({-i\epsilon p}/{\ehbar}\right)$ is the translation operator, and $B=|\psi(t_0)\rangle\langle\psi(t_0)|$ represents a projection operator onto an initial state, yielding the relation $C(t)=1-\langle \psi(t)|\exp\left({-i\epsilon p}/{\ehbar}\right)\psi(t)\rangle$. Our main results are summarized by the following relationships
\begin{align}\label{DCurrentTi}
\begin{aligned}
\langle p(t)\rangle=-\alpha\cos\left(\phi\right)Kt \:,
\end{aligned}
\end{align}
\begin{equation}\label{MEnergyTi}
\begin{aligned}
\langle p^2(t)\rangle=&\left[\left(\frac{3}{4}+2\alpha^2\right)K^2+\frac{2}{\pi}\alpha Kg\sin\left(\phi\right)+\frac{g^2}{2\pi^2}\right]t^2 +16\pi^2 \:,
\end{aligned}
\end{equation}
and
\begin{align}\label{OTOCsTim}
	\begin{aligned}
		C(t)\approx&\left(\frac{\epsilon}{4\pi}\right)^2\left\{\left[\frac{3}{4}+\alpha^2\left(2-\cos^2\phi\right)\right]K^2  \right.\\
		&\left.+\frac{2}{\pi}\alpha Kg\sin(\phi)+\frac{g^2}{2\pi^2}\right\}t^2\:,
	\end{aligned}
\end{align}
\begin{figure}[t]
\begin{center}
\includegraphics[width=8cm]{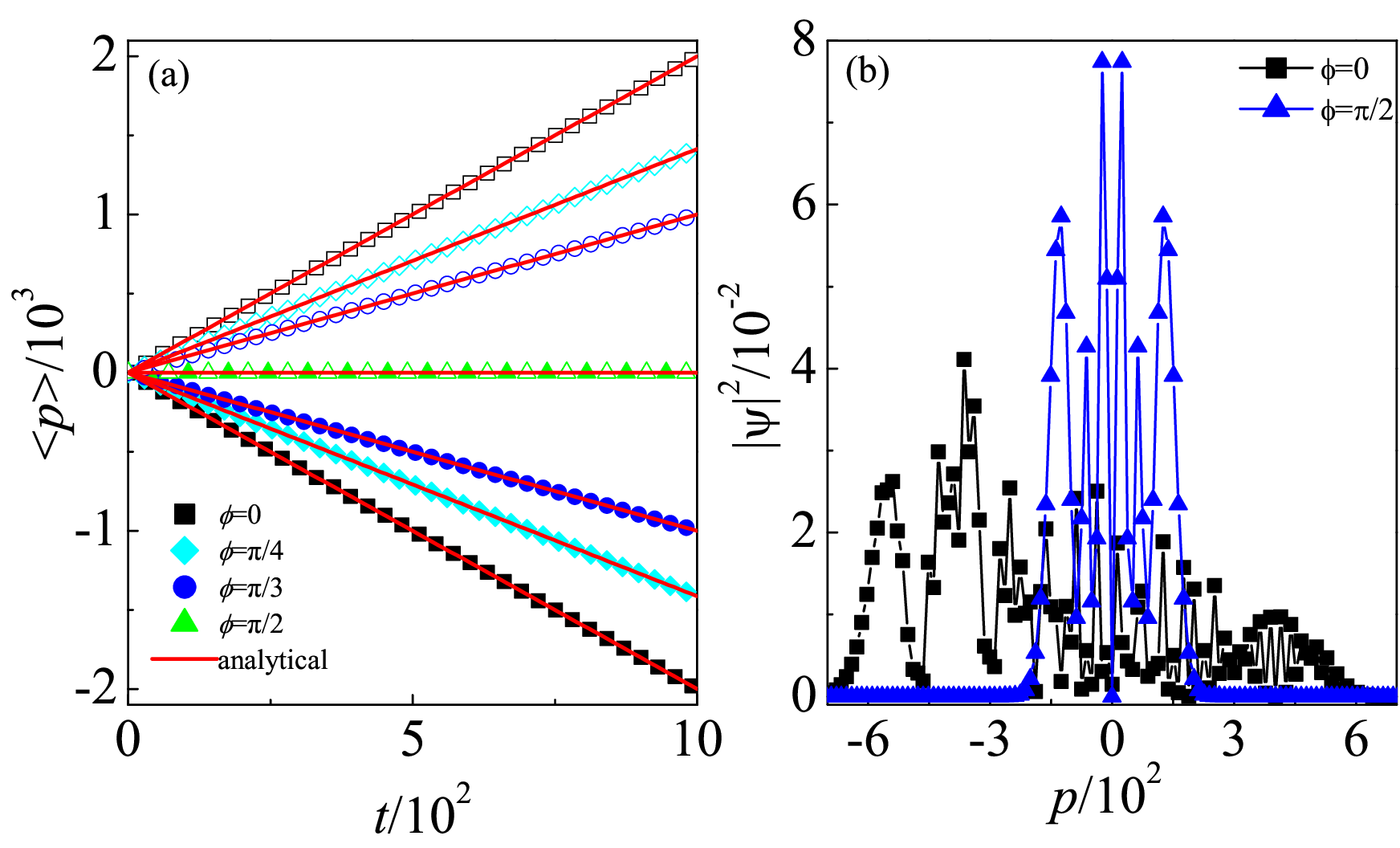}
\caption{(a) Time dependence of the $\langle p\rangle$ with $\phi=0$ (squares), $\pi/4$ (diamonds), $\pi/3$ (circles), and $\pi/2$ (triangles) for $\alpha=-2$ (empty symbols) and 2 (solid symbols).  Red lines indicate our theoretical prediction in Eq.~\eqref{DCurrentTi}. (b) Momentum distributions at the time $t=100$ with $\alpha=2$ for $\phi=0$ (squares) and $\pi/2$ (triangles). The parameters are $K=1$, $g=10$, and $\ehbar=4\pi$.\label{DCurrent}}
\end{center}
\end{figure}

Equation~\eqref{DCurrentTi} demonstrates the emergence of directed current in momentum space, which is unaffected by the self-interaction and can be engineered through the phase $\phi$ of the external potential. Our analytical prediction for the momentum current is confirmed by the numerical results shown in Fig.~\ref{DCurrent}(a). It is evident that the value of $\langle p \rangle$ remains zero for $\phi = 0$, while it increases linearly with time for a specific $\phi$ (e.g., $\phi=\pi/4$). In addition, the time dependece of $\langle p \rangle$ is governed by the parameter $\alpha$, which controls the amplitude of the ratchet potential, thereby enabling the adjustment of the directed current. We find that for $\phi = 0$, the momentum distribution is symmetric around $p = 0$, resulting in $\langle p \rangle = 0$ [see Fig.~\ref{DCurrent}(b)]. Interestingly, for $\phi = \pi/2$, a large portion of the momentum distribution is localized in the region with $p < 0$, leading to a negative average value, i.e., $\langle p \rangle < 0$. Therefore, the propagation of the quantum state in momentum space can be finely controlled by tuning the phase of the ratchet potential, which opens new opportunities for Floquet engineering in wavepacket dynamics~\cite{ZQLi23pra}. It is worth noting that under quantum resonance conditions, directed current can also emerge with a symmetric kicking potential if the initial state is antisymmetric~\cite{Zhaowl14PRE,Lundh06PRE}. However, in these cases, the directed current gradually disappears as the self-interaction increases~\cite{Zhaowl14PRE,Lundh06PRE}. In our system, the directed current is completely unaffected by the self-interaction, providing a robust method for realizing momentum current.

\begin{figure}[t]
\begin{center}
\includegraphics[width=8cm]{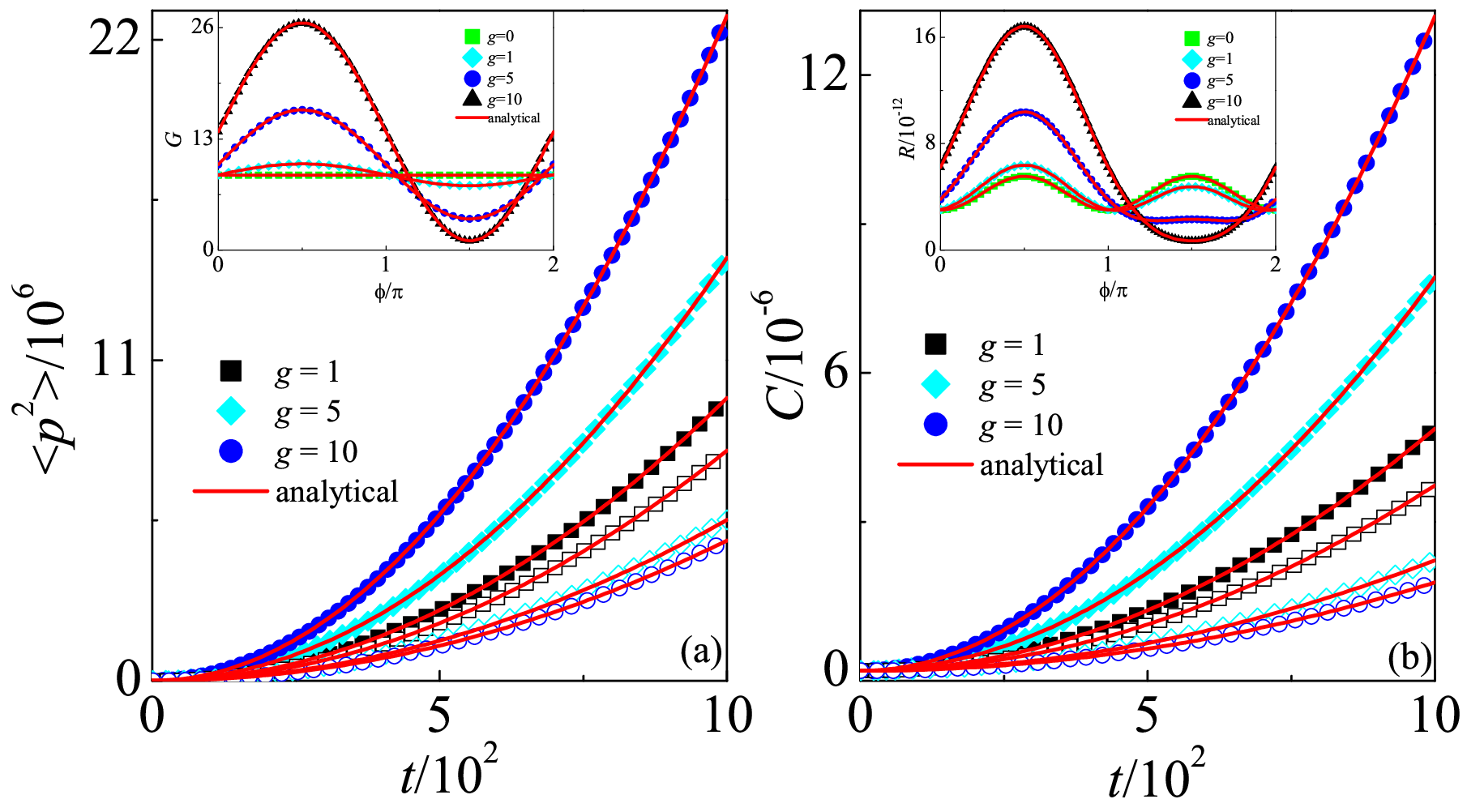}
\caption{Time dependence of $\langle p^2 \rangle$ (a) and $C$ (b) for $\phi = \pi/4$, $\alpha = 2$ (solid symbols) and $\alpha = -2$ (empty symbols), with $g = 1$ (squares), $g = 5$ (diamonds), and $g = 10$ (circles). Red lines in (a) and (b) indicate our theoretical predictions in Eqs.~\eqref{MEnergyTi} and~\eqref{OTOCsTim}. Inset in (a) and (b): Growth rate $G=\langle p^2(t)\rangle/{t^2}$ and $R=\langle C(t)\rangle/{t^2}$ versus $\phi$, with $\alpha = 2$, for $g=0$ (squares), 1 (diamonds), 5 (circles), and 10 (triangles). Red lines denotes our theoretical predictions. The value of translation parameter is $\epsilon=10^{-5}$. Other parameters are the same as in Fig.~\ref{DCurrent}. \label{Phsmodulation} }
\end{center}
\end{figure}

We also investigate numerically the time dependence of the mean energy $\langle p^2\rangle$ for different $\phi$. Figure~\ref{Phsmodulation}(a) shows that for a specific value of $g$ (e.g., $g=1$), the $\langle p^2 \rangle$ increases quadratically with time, in perfect agreement with our analytical prediction in Eq.~\eqref{MEnergyTi}. The dependence of the $\langle p^2 \rangle$ on system parameters $\alpha$, $K$, $g$, and $\phi$ is quantified by the growth rate $G=\langle p^2(t)\rangle/{t^2}=\left({3}/{4}+2\alpha^2\right)K^2+2\alpha Kg\sin\left(\phi\right)/\pi+{g^2}/{2\pi^2}$, which is confirmed by our numerical results in the inset of Fig.~\ref{Phsmodulation}(a). It is evident that the $G$ is independent of $\phi$ when $g=0$. However, for nonzero $g$, phase modulation in $G$ emerges, and the amplitude of this modulation increases linearly with $g$ [see the inset of Fig.~\ref{Phsmodulation}(a)]. It is worth noting that the phase modulation induced by self-interactions offers a new way to control the directed current by tuning the nonlinearity.

Our numerical results for the OTOCs demonstrate that for a specific value of $g$ [e.g., $g=1$ in Fig.~\ref{Phsmodulation}(b)], the $C$ follows a quadratic time dependence, perfectly matching Eq.~\eqref{OTOCsTim}. The growth rate, $R = C/t^2 = \epsilon^2 \left\{\left[3/4 + \alpha^2(2 - \cos^2\phi)\right]K^2 + 2\alpha Kg \sin(\phi)/\pi + g^2/2\pi^2\right\}/16\pi^2$, includes the term $2\alpha Kg \sin(\phi)/\pi$, which involves both $g$ and $\phi$, serving as evidence of self-interaction-induced phase modulation in quantum scrambling. Our numerical results for $R$ are in perfect agreement with the analytical prediction, as shown in the inset of Fig.~\ref{Phsmodulation}(b). We concentrate on the case with a very small translation parameter, i.e., $\epsilon \ll 1$. Based on the Taylor expansion $e^{-i\epsilon p} \approx 1 - i\epsilon p$, it is straightforward to obtain the relation
$C(t)\approx (\epsilon/\ehbar)^2 \left[\langle p^2(t)\rangle -\langle p(t)\rangle^2\right]$~\cite{Zhao24prr}, yielding the analytical prediction in Eq.~\eqref{OTOCsTim}. Thus, $C$ is proportional to the variance of a quantum state in momentum space, indicating an underlying correlation between quantum scrambling and energy diffusion. Our finding of quadratic growth in $C$ over time demonstrates ballistic energy diffusion, which can be controlled by tuning both the external potential parameters and the strength of self-interactions.

\section{Quantum nonresonance case} \label{QuanNReson}

\begin{figure}[t]
\begin{center}
\includegraphics[width=7.5cm]{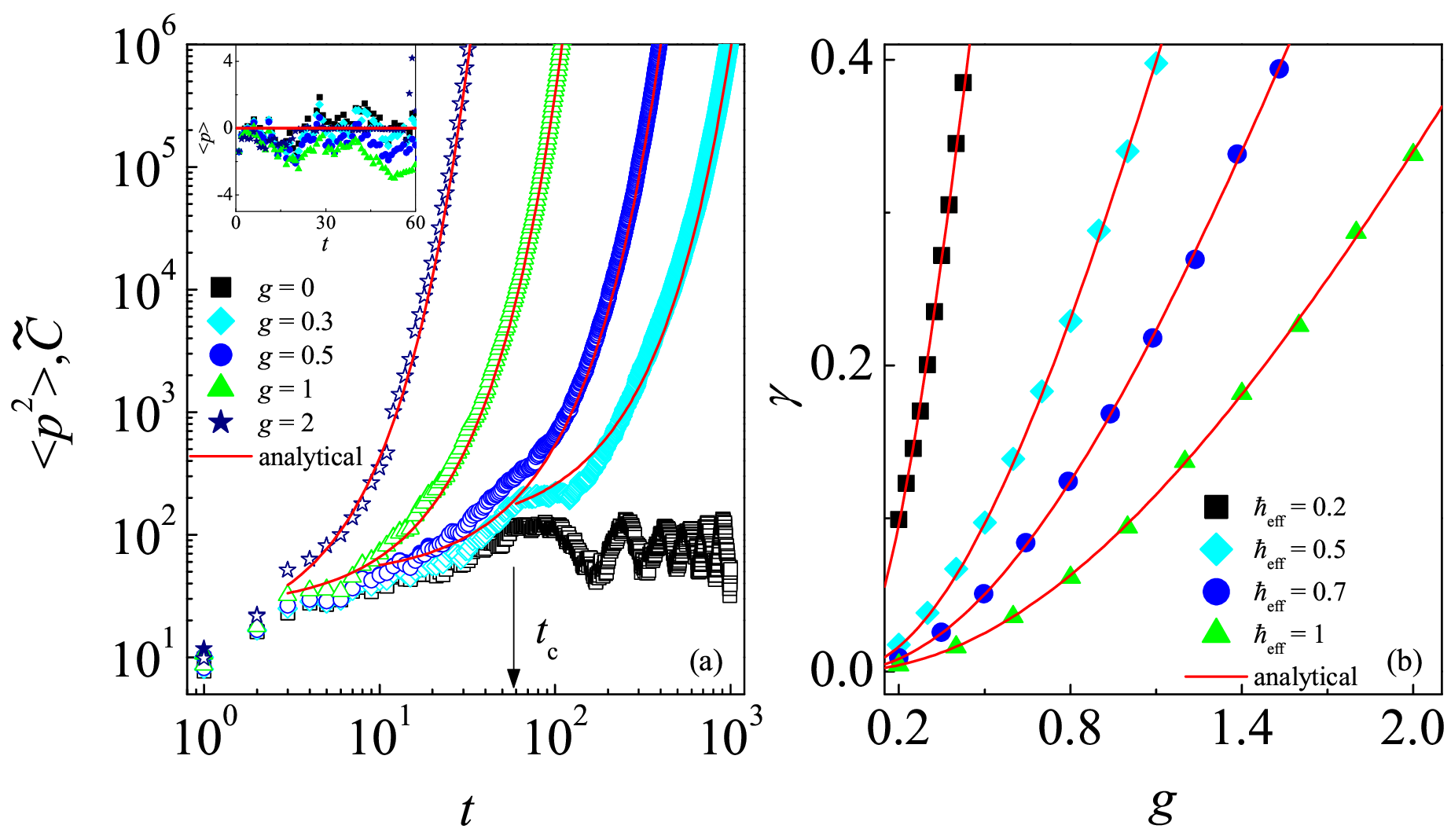}
\caption{(a) Time dependence of $\langle p^2 \rangle$ (solid symbols) and $\tilde{C}=C\hbar_{\text{eff}}^2/{\varepsilon}^2$ (empty symbols) for $\ehbar=1$, $\phi = \pi/4$, and $\alpha = 2$, with $g = 0$ (squares), $g = 0.3$ (diamonds), $g = 0.5$ (circles), $g = 1$ (triangles), and $g = 2$ (pentagrams). Red lines indicates our theoretical predictions in the form $\langle p^2\rangle \propto e^{\gamma t}$. Arrow marks the critical time $t_c$. Inset: Time dependence of $\langle p \rangle$ with $g = 0$ (squares), $g = 0.3$ (diamonds), $g = 0.5$ (circles), $g = 1$ (triangles), and $g = 2$ (pentagrams). Solid line indicates $\langle p \rangle=0$. (b): Growth rate $\gamma$ versus $g$, with $\ehbar = 0.2$ (squares), 0.5 (diamonds), 0.7 (circles), and 1 (triangles). Red lines denotes our theoretical predictions $\gamma \approx \ln\left[1 + \left({g}/{\pi \ehbar}\right)^2\right]$. Other parameters are the same as in Fig.~\ref{DCurrent}. \label{Expdiffusion}}
\end{center}
\end{figure}

It is well known that in the QKR model, rich phenomena such as dynamical localization~\cite{Chirikov91} and the quantum boomerang effect~\cite{Noronha22prb,Tessieri21pra,Sajjad22prx} occur in the quantum nonresonance regime, i.e., when $\ehbar\neq 4\pi r/s$, where $r$ and $s$ are coprime integers. Therefore, we further investigate the time evolution of momentum current, mean energy and quantum scrambling in quantum non-resonance condition. We find that for different $g$, the $\langle p \rangle$ fluctuates around zero over time, demonstrating the disappearance of the directed current. The underlying physics results from momentum-space disorder generated by the pseudorandom nature of the free evolution operator~\cite{Grempel84pra}. The mean energy clearly exhibits dynamical localization during time evolution when $g = 0$. Interestingly, for nonzero values of $g$ (e.g., $g = 0.3$ in Fig.~\eqref{Expdiffusion}(a)), $\langle p^2 \rangle$ follows the behavior of $g = 0$ for a finite period, i.e., $t < t_c$, then grows exponentially, i.e., $\langle p^2 \rangle \propto e^{\gamma t}$ when $t > t_c$. Notably, the mean energy increases significantly faster for larger $g$, while the critical time $t_c$ decreases as $g$ increases. We also numerically investigate the growth rate $\gamma$ for different values of $g$. The dependence of $\gamma$ on $g$ perfectly matches the analytical expression $\gamma \approx \ln\left[1 + \left({g}/{\pi \ehbar}\right)^2\right]$, which arises from the exponential diffusion in the generalized QKR model (see Appendix). Since $\langle p \rangle$ is negligibly small compared to $\langle p^2 \rangle$, we derive the relation $C(t) \approx \left({\epsilon}/{\ehbar}\right)^2 \langle p^2(t) \rangle$, which is verified by our numerical results shown in Fig.~\eqref{Expdiffusion}(a).

\begin{figure}[t]
\begin{center}
\includegraphics[width=8cm]{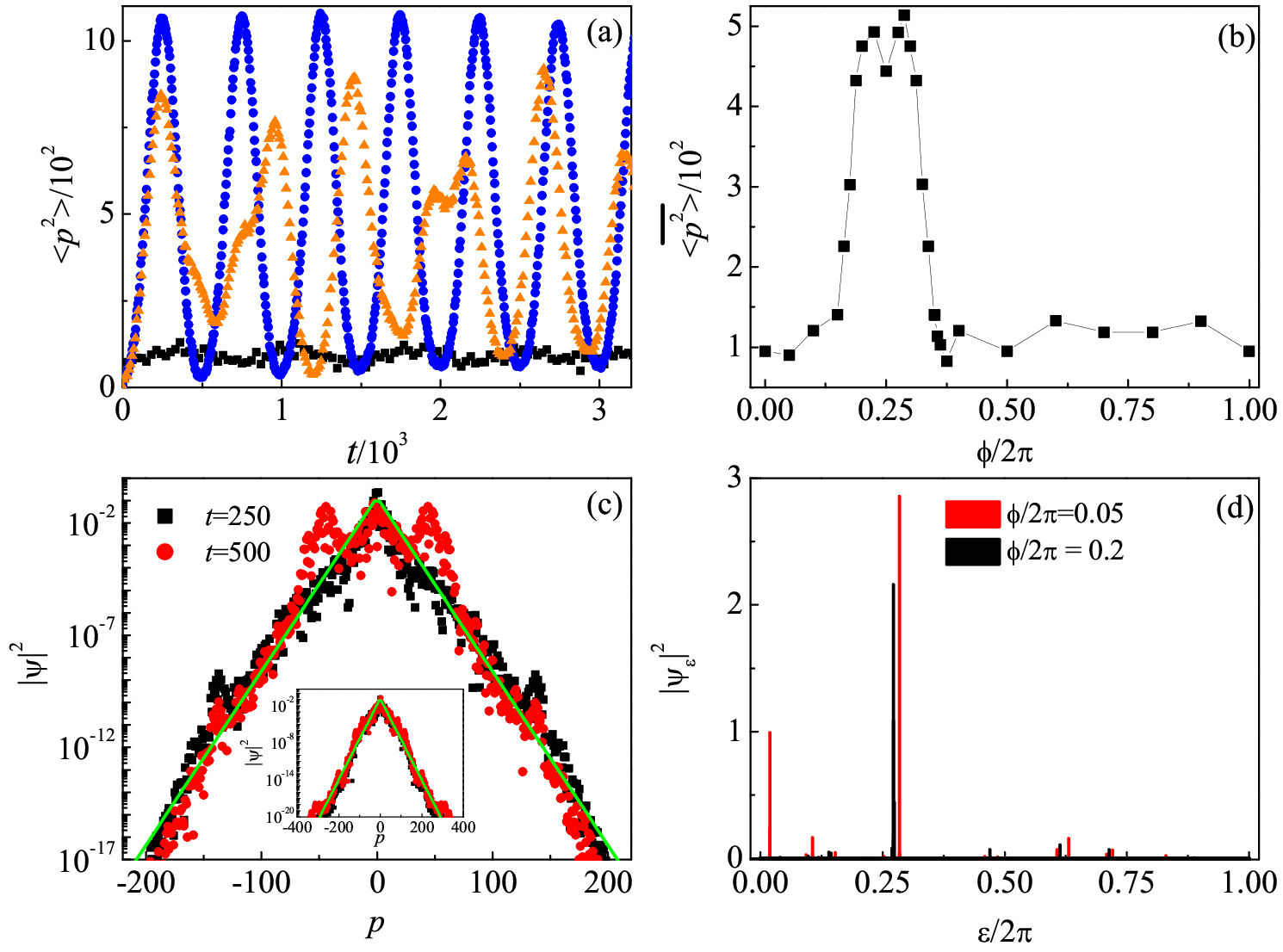}
\caption{(a) Time dependence of $\langle p^2 \rangle$ with $\phi/2\pi=0.05$ (squares), $0.2$ (circles), and $0.25$ (triangles). (b) The $\overline{\langle p \rangle}$ versus $\phi$. (c) Momentum distributions with $\phi/2\pi=0.2$ for $t=250$ (squares) and $500$ (circles). Green line indicates the fitting function of the form $|\psi(p)|\propto\exp(-|p|/\xi)$ with $\xi\approx 5.6$. Inset: Same as in the main plot but for $\phi/2\pi=0.05$. Green line represents the exponential fitting, i.e., $|\psi(p)|\propto\exp(-|p|/\xi)$ with $\xi\approx 6.6$. (d) The $|\psi_{\lambda}|^2$ versus $\lambda$ for $\phi/2\pi=0.05$ (red bar) and 0.2 (black bar). The parameters are $K=1$, $\alpha = 2$, $\ehbar=1$, and $g = 0$. \label{dynlocal}}
\end{center}
\end{figure}

We further investigate the dynamical localization of energy diffusion with $g=0$ for different values of $\phi$. Our results show that for $\phi/2\pi = 0.05$, $\langle p^2 \rangle$ rapidly saturates over time [see Fig.\ref{dynlocal}(a)]. The quantum states are exponentially localized in momentum space, i.e., $|\psi(p)|^2 \propto \exp(-|p|/\xi)$, with a constant localization length $\xi$ [see the inset in Fig.\ref{dynlocal}(c)]. Interestingly, for $\phi/2\pi = 0.2$, the mean energy oscillates periodically over time, indicating a new form of dynamical localization induced by the ratchet potential. The comparison between the momentum distributions at the maximum and minimum of $\langle p^2 \rangle$, corresponding to $t = 250$ and 500, shows that both distributions exhibit exponentially localized tails, while $|\psi(p)|^2$ at $t = 250$ displays two significant peaks around $|p|\approx 50$ [see Fig.~\ref{dynlocal}(c).], leading to the maximum mean energy. For a slightly larger $\phi$, i.e., $\phi/2\pi = 0.25$, the time evolution of $\langle p^2 \rangle$ shows clear oscillations, though without a perfect period. To quantify the saturation value of these oscillations we numerically investigate the time-averaged value $\overline{\langle p^2\rangle}$ for different values of $\phi$~\cite{NoteSaturation}. Our results show that $\overline{\langle p^2\rangle}$ remains approximately 100 as $\phi$ varies, except in the region $0.125 \lesssim \phi/2\pi \lesssim 0.375$, where it increases to about 450. In this region, $\langle p^2\rangle$ exhibits significant oscillations over time. Our investigations demonstrate that, in the quantum nonresonance regime, the delta-kicking ratchet potential offers an opportunity to engineer the behavior of dynamical localization, even though it does not produce a directed current.

In order to reveal the underlying mechanism for different behaviors of dynamical localization, we numerically investigate the distributions of quantum states in quasi-eigenenergy space. The eigenvalue equation of the Floquet operator is given by $U|\varphi_{\varepsilon}\rangle = e^{-i\varepsilon}|\varphi_{\varepsilon}\rangle$. With this complete basis, an initial state can be expressed as $|\psi(t_0)\rangle = \sum_{\varepsilon}\psi_{\varepsilon} |\varphi_{\varepsilon}\rangle$. After the $n$th kicks, the quantum state is given by $|\psi(t_n)\rangle = \sum_{\varepsilon}\psi_{\varepsilon} e^{-i\varepsilon t_n}|\varphi_{\varepsilon}\rangle$. This leads to the auto-correlation function $A(t_n) = \langle\psi(t_0)|\psi(t_n)\rangle = \sum\limits_{\varepsilon}|\psi_\varepsilon|^2 e^{-i\varepsilon t_n}$. It is apparent that the Fourier components of $A(t_n)$ indicate the probability density distribution, i.e., $|\psi_\varepsilon|^2$ in the quasi-eigenenergy representation~\cite{Grempel84pra,JWang10pre}. We numerically investigate $|\psi_\varepsilon|^2$ for different $\phi$. Figure~\ref{dynlocal}(d) shows that for $\phi/2\pi = 0.05$, two significant peaks in $|\psi_\varepsilon|^2$ govern the expansion of the quantum state $|\psi(t_n)\rangle$. The saturation behavior of $\langle p^2 \rangle$ likely results from the interference effects of the quasi-eigenstates corresponding to these two peaks. Interestingly, for $\phi/2\pi = 0.2$, the probability density distribution $|\psi_\varepsilon|^2$ has only one significant peak, yielding a single-frequency oscillation of the mean energy.

\section{Conclusion and discussions}\label{Sum}

In this work, we investigate both analytically and numerically the interplay between self-interaction and phase on the dynamics of $\langle p \rangle$, $\langle p^2 \rangle$, and $C$ in a kicked ratchet rotor model. The ratchet effects lead to the linear growth of $\langle p \rangle$, indicating the emergence of directed current, with the acceleration rate $\langle p \rangle/t$ governed by $\phi$. Both $\langle p^2 \rangle$ and $C$ are quadratic functions of time, where their growth rates contain a term proportional to the product of $g$ and $\sin(\phi)$, demonstrating nonlinearity-induced phase modulation. In the quantum nonresonance case, the directed current disappears, resulting in the proportional equivalence between $C$ and $\langle p^2 \rangle$. The mechanism of dynamical localization suppresses both $C$ and $\langle p^2 \rangle$ for $t < t_c$, beyond which they both increase exponentially with time, following the classical diffusion of a GKR model. Interestingly, for certain values of $\phi$, $\langle p^2 \rangle$ exhibits periodic oscillations over time, which is distinct from the conventional saturation of $\langle p^2 \rangle$ in dynamical localization. The underlying quasi-eigenenergy spectrum, obtained from the Fourier transform of the autocorrelation function $A(t)$, shows a significant peak, which accounts for the single-frequency oscillations of the mean energy. Our findings suggest new possibilities for engineering directed current, energy diffusion, and quantum scrambling in Floquet systems by adjusting nonlinearity and phase, shedding light on quantum control in chaotic systems~\cite{Suraj24pla}.

\section*{ACKNOWLEDGMENTS}
This work is supported by the National Natural Science Foundation of China (Grant No. 12365002 and 12065009), the Science and Technology Planning Project of Jiangxi province (Grant No. 20224ACB201006 and 20224BAB201023).

\appendix

\section{Classical limit of the QKR with self-interaction}

The dimensionless Hamiltonian of the QKR model with self-interaction reads
\begin{equation}
\begin{aligned}\label{HamilAPX}
{\rm H} = \frac{p^2}{2} + \left[g|\psi(\theta,t)|^2 + V_K(\theta)\right]\sum\limits_n\delta(t - t_n) \:,
\end{aligned}
\end{equation}
with $V_K(\theta) = K[\cos(\theta) + i\sin(\theta)]$. For brevity, we define the nonlinear kicking potential as
\begin{align}
\begin{aligned}\label{NKPotenAPX}
V(\theta,t) = g|\psi(\theta,t)|^2 + V_K(\theta) \:.
\end{aligned}
\end{align}
For periodic functions, i.e., $V(\theta,t) = V(\theta + 2\pi,t)$, the Fourier expansion takes the form
$V(\theta,t)=\sum\limits_{n=-\infty}^{+\infty}V_n(t)e^{in\theta}/{\sqrt{2\pi}}$
where the complex components are $V_n(t) = V_n^r(t) + iV_n^i(t)$. Straightforward derivation yields the expression
\begin{align}
\begin{aligned}\label{FExpasion}
V(\theta,t)=&\frac{V_0^r(t)}{\sqrt{2\pi}}+\sum_{n=1}^{\infty}\left[K_n^r(t)\cos(n\theta)-K_n^i(t)\sin(n\theta)\right]\:,
\end{aligned}
\end{align}
where $K_n^r(t)=[V_n^r(t)+V_{-n}^r(t)]/{\sqrt{2\pi}}$ and $K_n^i(t)=[V_n^i(t)+V_{-n}^i(t)]/{\sqrt{2\pi}}$.
Taking Eq.~\eqref{FExpasion} into Eq.~\eqref{HamilAPX} yields
\begin{equation}
\begin{aligned}\label{HamilAPX2}
{\rm H} \approx \frac{p^2}{2} + \sum_{n=1}^{\infty}\left[K_n^r(t)\cos(n\theta) - K_n^i(t)\sin(n\theta)\right]\delta(t - t_n)\:,
\end{aligned}
\end{equation}
where the term ${V_0^r(t)}/{\sqrt{2\pi}}$ in $V(\theta,t)$ is dropped because it is independent of $\theta$, and thus only contributes trivial overall phases to the time-evolving state after each kick.

\begin{figure}[t]
\begin{center}
\includegraphics[width=8.0cm]{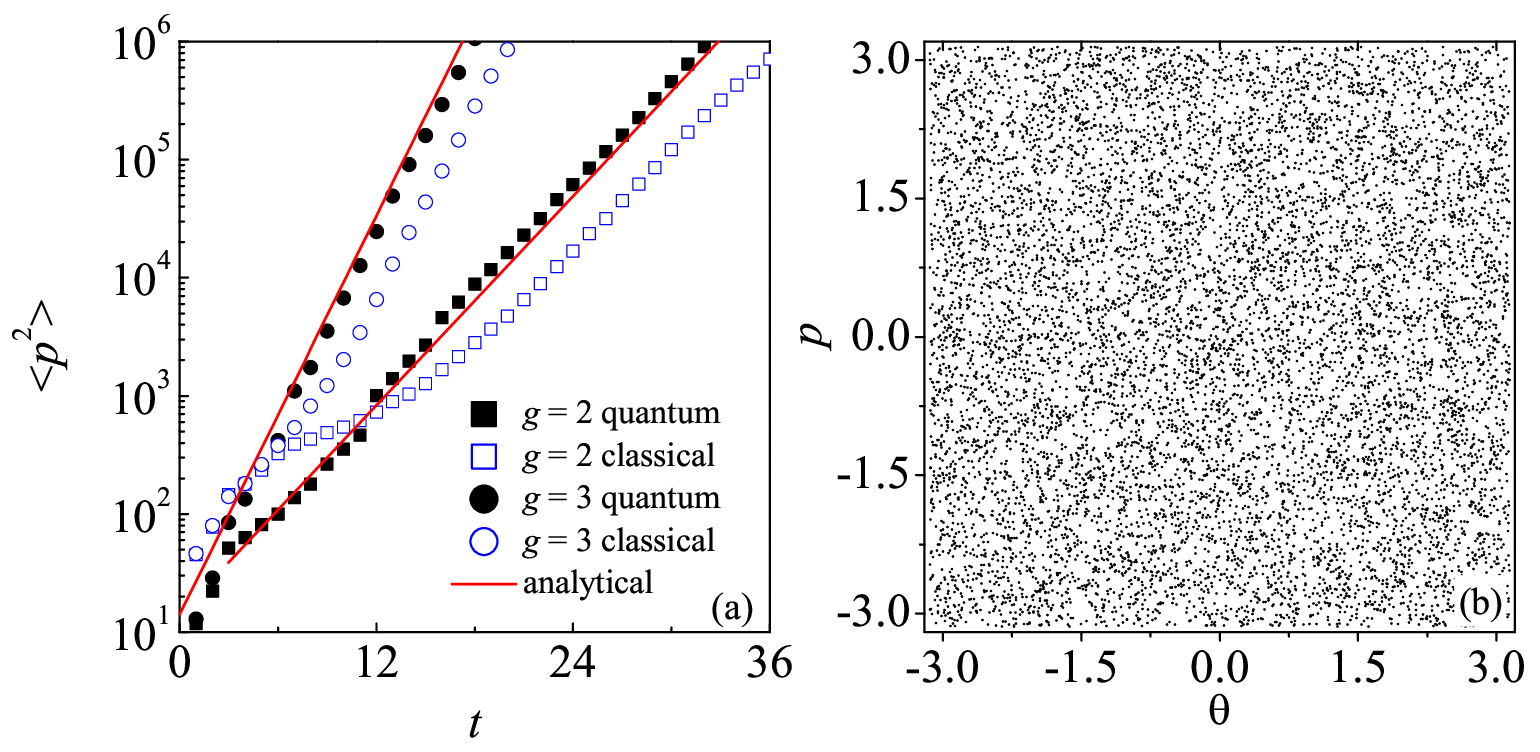}
\caption{(a) Comparison in mean energy between the classical GKR (empty symbols) and the quantum results (solid symbols) with $g$=2 (squares) and 3 (circles). Red lines indicate analytical prediction $\langle p^2 \rangle \propto e^{\gamma t}$. (b) Phase space portrait of the classical GKR model for an ensemble of $N=10^4$ trajectories with $t=10$ and $g=3$. The parameters are $K=1$, $\alpha = 2$, $\ehbar=1$, and $\phi = \pi/4$. \label{QCC}}
\end{center}
\end{figure}

The Hamiltonian represents a generalized kick rotor (GKR) model, with the kick strengths $K_n^r(t)$ and $K_n^i(t)$ dependent on the quantum state of the kicked Gross-Pitaevskii system~\cite{wlzhao16pra,wlzhao19jpa}. The corresponding classical mapping equations for Eq.~\eqref{HamilAPX2} are given by
\begin{equation}
\begin{aligned}\label{CLMapping}
p(t+1) - p(t) &= \sum_{n=1}^{\infty}\left[nK_n^r(t)\sin(n\theta) + nK_n^i(t)\cos(n\theta)\right]\;,\\
\theta(t+1) - \theta(t) &= p(t+1)\;,
\end{aligned}
\end{equation}
where $p$ and $\theta$ indicate the classical momentum and coordinate, respectively. These equations enable the investigation of time evolution in classical trajectories, providing insights into the underlying classical dynamics~\cite{wlzhao16pra,wlzhao19jpa}. We numerically calculate the time evolution of the ensemble-averaged classical mean energy $\langle p^2 \rangle_{\text{cl}}$, where $\langle \cdots \rangle$ denotes the average over classical trajectories. In the simulations, the initial values of the trajectories are set with $\theta$ uniformly distributed over $[0, 2\pi]$ and $p = 0$. Our results demonstrate that after long-term evolution, the classical mean energy $\langle p^2 \rangle_{\text{cl}}$ is approximately proportional to the quantum mean energy $\langle p^2 \rangle$, and both exhibit exponential growth over time, i.e., $\langle p^2 \rangle_{\text{cl}} \propto e^{\gamma t}$ [see Fig.~\ref{QCC}(a)]. This behavior is rooted in the chaotic dynamics, as indicated by the fully chaotic sea in the classical phase space [see Fig.~\ref{QCC}(b)]. Based on classically chaotic diffusion, we can obtain the dependence of the growth rate $\gamma$ on both $g$ and $\ehbar$
\begin{equation}
\gamma \approx \ln\left[1+\left(\frac{g}{\pi \ehbar}\right)^2\right]\;.
\end{equation}
Detailed derivations can be found in our previous work, i.e., Refs.~\cite{wlzhao16pra,wlzhao19jpa}.


\begin{thebibliography}{*}

\bibitem{SMU24prl}
S. Mu, J. B. Gong, and G. Lemari{\' e}, Kardar-Parisi-Zhang Physics in the Density Fluctuations of Localized Two-Dimensional Wave Packets, Phys. Rev. Lett. {\bf 132}, 046301 (2024).

\bibitem{Graham05jpa}
B. Mieck and R. Graham, Bose-Einstein condensate of kicked rotators with time-dependent interaction, J. Phys. A: Math. Gen. {\bf 38}, L139 (2005).

\bibitem{Guarneri17pre}
I. Guarneri, Gross-Pitaevski map as a chaotic dynamical system, Phys. Rev. E {\bf 95}, 032206 (2017).

\bibitem{WLZhao21prb}
W. L. Zhao, Y. Hu, Z. Li, and Q. Wang, Super-exponential growth of Out-of-time-ordered correlators, \prb {\bf 103}, 184311 (2021).


\bibitem{wlzhao16pra}
W. L. Zhao, J. Gong, W. G. Wang, G. Casati, J. Liu, and L. B. Fu, Exponential wave-packet spreading via self-interaction time modulation, Phys. Rev. A {\bf 94}, 053631 (2016).

\bibitem{wlzhao19jpa}
W. Zhao, J. Z. Wang, and W. Wang, Quantum-classical correspondence
in a nonlinear Gross-Pitaevski system, J. Phys. A: Math. Theor. {\bf 52}, 305101 (2019).

\bibitem{SMU22epl}
S. Mu, N. Mac{\' e}, J. Gong, C. Miniatura, G. Lemari{\' e}, and M. Albert, Superfluidity vs. prethermalisation in a nonlinear Floquet system, EPL, {\bf 140}, 50001 (2022).

\bibitem{Haldar23epl}
P. Haldar, S. Mu, B. Georgeot, J. B. Gong, C. Miniatura, and G. Lemari{\' e}, Rayleigh-Jeans, prethermalization and wave condensation in a nonlinear disordered Floquet system, EPL, {\bf 144}, 63001 (2023).

\bibitem{Banerjee24arx}
T. Banerjee and K. Sengupta, Emergent symmetries in prethermal phases of periodically driven quantum systems, arXiv:2407.20764.


\bibitem{Meng22pra}
L. Z. Meng, S. W. Guan, and L. C. Zhao, Negative mass effects of a spin soliton in Bose-Einstein condensates,
Phys. Rev. A {\bf 105}, 013303 (2022).


\bibitem{Bomantara17prb}
R. W. Bomantara, W. L. Zhao, L. W. Zhou, and J. B. Gong, Nonlinear Dirac cones, Phys. Rev. B {\bf 96}, 121406(R) (2017).


\bibitem{Tuloup20prb}
T. Tuloup, R. W. Bomantara, C. H. Lee, and J. B. Gong, Nonlinearity induced topological physics in momentum space and real space, Phys. Rev. B {\bf 102}, 115411 (2020).

\bibitem{Goussev18pra}
A. Goussev, P. Reck, F. Moser, A. Moro, C. Gorini, and K. Richter, Overcoming dispersive spreading of quantum wave packets via periodic nonlinear kicking, Phys. Rev. A {\bf 98}, 013620 (2018).

\bibitem{WLZhao23arx}
W. L. Zhao and J. Liu, Superexponential behaviors of out-of-time ordered correlators and Loschmidt echo in a non-Hermitian interacting system, arXiv:2305.1215.

\bibitem{Duval22pra}
C. Duval, Dominique Delande, and Nicolas Cherroret, Subdiffusion in wave packets with periodically kicked interactions,
Phys. Rev. A {\bf 105}, 033309 (2022).


\bibitem{JBGong02prl}
J. B. Gong and P. Brumer, Phase Control of Nonadiabaticity-Induced Quantum Chaos in an Optical Lattice, \prl {\bf 88}, 203001 (2002).

\bibitem{JBGong01prl}
J. B. Gong and P. Brumer, Coherent Control of Quantum Chaotic Diffusion, \prl {\bf 86}, 1741 (2001).

\bibitem{Bitter17prl}
M. Bitter and V. Milner, Experimental Demonstration of Coherent Control in Quantum Chaotic Systems, \prl {\bf 118}, 034101 (2017).

\bibitem{Sentef20prr}
M. A. Sentef, J. J. Li, F. K{\" u}nzel, and M. Eckstein, Quantum to classical crossover of Floquet engineering in correlated quantum systems, Phys. Rev. Research {\bf 2}, 033033 (2020).

\bibitem{Bai20pra}
S. Y. Bai and J. H. An, Floquet engineering to reactivate a dissipative quantum battery, \pra {\bf 102}, 060201(R) (2020).

\bibitem{Downing23}
C. A. Downing and M. S. Ukhtary, A quantum battery with quadratic driving, Commun. Phys. {\bf 6}, 322 (2023).

\bibitem{Kenfack08prl}
A. Kenfack, J. B. Gong, and A. K. Pattanayak, Controlling the Ratchet Effect for Cold Atoms, \prl {\bf 100}, 044104 (2008).

\bibitem{Hainaut18pra}
C. Hainaut, A. Ran{\c c}on, J. Cl{\' e}ment, J. C. Garreau, P. Szriftgiser, R. Chicireanu, and D. Delande, Ratchet effect in the quantum kicked rotor and its destruction by dynamical localization
\pra {\bf 97}, 061601(R) (2018).

\bibitem{ZQLi23pra}
Z. Q. Li, X. X. Hu, J. P. Xiao, Y. J. Chen, and X. B. Luo, Ratchet current in a $\cal{PT}$-symmetric Floquet quantum system with symmetric sinusoidal driving, \pra {\bf 108}, 052211 (2023).

\bibitem{Poletti07pra}
D. Poletti, G. Benenti, G. Casati, and B. W. Li, Interaction-induced quantum ratchet in a Bose-Einstein condensate, \pra {\bf 76}, 023421 (2007).

\bibitem{FXChen24arx}
F. X. Chen and P. Fang, System Symmetry and the Classification of Out-of-Time-Ordered Correlator Dynamics in Quantum Chaos, arXiv:2410.04712.


\bibitem{Meier19pra}
E. J. Meier, J. Ang'ong'a, F. Alex An, and B.  Gadway, Exploring quantum signatures of chaos on a Floquet synthetic lattice, \pra {\bf 100}, 013623 (2019).

\bibitem{JHWang22prr}
J. H. Wang {\it et al.,} Information scrambling dynamics in a fully controllable quantum simulator, Phys. Rev. Research {\bf 4}, 043141 (2022).

\bibitem{JHarris22prl}
J. Harris, B. Yan, and N. A. Sinitsyn, Benchmarking Information Scrambling, \prl {\bf129}, 050602 (2022).


\bibitem{Zhang15SB}
C. Zhang, C. F. Li, and G. C. Guo, Experimental demonstration of photonic quantum ratchet, Sci Bull {\bf 60}, 249 (2015).

\bibitem{Casati89prl}
G. Casati,  I. Guarneri, and D. L. Shepelyansky, Anderson transition in a onedimensional system with three incommensurate frequencies, \prl {\bf 62}, 345 (1989).

\bibitem{Shepelyansky87}
D. L. Shepelyansky, Localization of diffusive excitation in multi-level systems, Physica D {\bf 28}, 103 (1987).



\bibitem{Tian11prl}
C. S. Tian, A. Altland, and M. Garst, Theory of the Anderson Transition in the Quasiperiodic Kicked Rotor, Phys. Rev. Lett. {\bf 107}, 074101 (2011).


\bibitem{Lopez12prl}
M. Lopez, J. F. Cl{\' e}ment, P. Szriftgiser, J. C. Garreau, and D. Delande, Experimental Test of Universality of the Anderson Transition, \prl {\bf 108}, 095701 (2012).

\bibitem{Cherroret14prl}
N. Cherroret, B. Vermersch, J. C. Garreau, and D. Delande, How Nonlinear Interactions Challenge the Three-Dimensional Anderson Transition,
Phys. Rev. Lett. {\bf 112}, 170603 (2014).


\bibitem{Lopez13}
M. Lopez, J. F. Cl{\' e}ment, G. Lemari{\' e}, D. Delande, P. Szriftgiser, and J. C. Garreau, Phase diagram of the anisotropic Anderson transition with the atomic kicked rotor: theory and experiment, New J. Phys. {\bf 15}, 065013 (2013).

\bibitem{Manai15prl}
I. Manai, J. Cl{\' e}ment, R. Chicireanu, C.  Hainaut, J. C. Garreau, P. Szriftgiser, and D.  Delande, Experimental Observation of Two-Dimensional Anderson Localization with the Atomic Kicked Rotor, \prl {\bf 115}, 240603 (2015).



\bibitem{Hainaut18prl}
C. Hainaut, P. Fang, A. Ran{\c c}on, J. F. Cl{\' e}ment, Pascal Szriftgiser, J. C Garreau, C. S. Tian, and R. Chicireanu, Experimental Observation of a Time-Driven Phase Transition in Quantum Chaos, \prl {\bf 121}, 134101 (2018).

\bibitem{Chen14prl}
Y. Chen and C. S. Tian, Planck's Quantum-Driven Integer Quantum Hall Effect in Chaos, \prl {\bf 113}, 216802 (2014).

\bibitem{Derek12prl}
Derek Y. H. Ho, and J. B. Gong, Quantized adiabatic transport in momentum space, \prl {\bf 109}, 010601 (2012).


\bibitem{Sharabi18}
Y. Sharabi, H. Sheinfux, Y. Sagi, G. Eisenstein, and M. Segev, Self-induced diffusion in disordered nonlinear photonic media, \prl {\bf 121}, 233901 (2018).

\bibitem{Prange89}
R. E. Prange and S. Fishman, Experimental realizations of kicked quantum chaotic systems, \prl {\bf 63}, 704 (1989).

\bibitem{Fischer00pre}
B. Fischer, A. Rosen, A. Bekker, and S. Fishman, Experimental observation of localization in the spatial frequency domain of a kicked optical system, Phys. Rev. E {\bf 61}, 4694(R) (2000).

\bibitem{Rosen00JOS}
A. Rosen, B. Fischer, A. Bekker, and S. Fishman, Optical kicked
system exhibiting localization in the spatial frequency domain, J. Opt. Soc. Am. B {\bf 17}, 1579 (2000).

\bibitem{Summy16PRA}
G. Summy and S. Wimberger, Quantum random walk of a Bose-Einstein condensate in momentum space, Phys. Rev. A {\bf 93}, 023638 (2016).



\bibitem{Dadras19}
S. Dadras, A. Gresch, C. Groiseau, S. Wimberger, and G. S. Summy, Experimental realization of a momentum-space quantum walk, Phys. Rev. A {\bf 99}, 043617 (2019).

\bibitem{Dadras18PRL}
S. Dadras, A. Gresch, C. Groiseau, S. Wimberger, and G. S. Summy, Quantum Walk in Momentum Space with a Bose-Einstein Condensate, Phys. Rev. Lett. {\bf 121}, 070402 (2018).


\bibitem{HLWang13PRE}
H. L. Wang, J. Wang, I. Guarneri, G. Casati, and J. B. Gong, Exponential quantum spreading in a class of kicked rotor systems near high-order resonances, Phys. Rev. E {\bf 88}, 052919 (2013).

\bibitem{Bomantara16PRE}
R. W. Bomantara, G. N. Raghava, L. W. Zhou, and J. B. Gong, Phys. Rev. E {\bf 93}, 022209 (2016).

\bibitem{Zhou14epjb}
L. Zhou, H. Wang, D. Y. H. Ho, and J. Gong, Aspects of Floquet bands and topological phase transitions in a continuously driven superlattice, Eur. Phys. J. B {\bf 87}, 204 (2014).

\bibitem{JWang08PRA}
J. Wang and J. B. Gong, Proposal of a cold-atom realization of quantum maps with Hofstadter's butterfly spectrum, Phys. Rev. A {\bf 77}, 031405(R) (2008).


\bibitem{Zhoulw14}
L. W. Zhou and J. X. Pan, Non-Hermitian Floquet topological phases in the double-kicked rotor, Phys. Rev. A {\bf 100}, 053608 (2019).

\bibitem{LWZhou23arx}
L. W. Zhou, D.-J. Zhang, Non-Hermitian Floquet Topological Matter---A Review, Entropy {\bf 25}, 1401 (2023).

\bibitem{Lundh05PRL}
E. Lundh and M Wallin, Ratchet effect for cold atoms in an optical lattice, \prl {\bf 94}, 110603 (2005).

\bibitem{Zhaowl14PRE}
W. L. Zhao, L. B. Fu, and J. Liu, Nonlinearity effects on the directed momentum current, Phys. Rev. E {\bf 90}, 022907 (2014).

\bibitem{Lundh06PRE}
E. Lundh, Directed transport and Floquet analysis for a periodically kicked wave packet at a quantum resonance, Phys. Rev. E {\bf 74}, 016212 (2006).

\bibitem{ZQi23}
Z. Qi, T. Scaffidi, and X. Cao, Surprises in the deep Hilbert space of all-to-all systems: From superexponential scrambling to slow entanglement growth, \prb {\bf 108}, 054301 (2023).

\bibitem{XDHu23}
X. D. Hu, T. Luo, and D. B. Zhang, Quantum algorithm for evaluating operator size with Bell measurements, \pra {\bf 107}, 022407 (2023).

\bibitem{JWang21pre}
J. Wang, G. Benenti, G. Casati, and W. G. Wang, Quantum chaos and the correspondence principle, \pre {\bf 103}, L030201 (2021).

\bibitem{Gribben24arx}
D. Gribben, J. Marino, and S. P. Kelly, Markovian to non-Markovian phase transition in the operator dynamics of a mobile impurity, arXiv:2401.17066.

\bibitem{Sharma24arx}
K. Sharma, H. Sahu, and S. Mukerjee, Quantum chaos in PT symmetric quantum systems, arXiv:2401.07215.

\bibitem{Rozenbaum17prl}
E. B. Rozenbaum, S. Ganeshan, and V. Galitski, Lyapunov Exponent and Out-of-Time-Ordered Correlator's Growth Rate in a Chaotic System, \prl {\bf 118}, 086801 (2017).

\bibitem{Garc24prd}
A. M. Garc{\' ia}-Garc{\' ia}a, J. J. M. Verbaarschot, and J.-P. Zheng, Lyapunov exponent as a signature of dissipative many-body quantum chaos, Phys. Rev. D {\bf 110}, 086010 (2024).




\bibitem{Pappalardi22SP}
S. Pappalardi and J. Kurchan, Low temperature quantum bounds on simple models, SciPost Phys. {\bf 13}, 006 (2022).

\bibitem{Khalouf24pra}
J. Khalouf-Rivera, Q. Wang, L. F. Santos, J.-E. Garc{\' i}a-Ramos, M Carvajal, and F. P{\' e}rez-Bernal, Degeneracy in excited-state quantum phase transitions of two-level bosonic models and its influence on system dynamics, Phys. Rev. A {\bf 109}, 062219 (2024).

\bibitem{SYWang23arx}
S. Y. Wang, S. B. Chen, J. L. Jing, J. C. Wang, and H. Fan, Quantum collapse and exponential growth of out-of-time-ordered correlator in anisotropic quantum Rabi model, arXiv:2305.17495.

\bibitem{Zonnios22}
M. Zonnios, J. Levinsen, M. M. Parish, F. A. Pollock, and K. Modi, Signatures of Quantum Chaos in an Out-of-Time-Order Tensor, \prl {\bf 128}, 150601 (2022).

\bibitem{Hashimoto2017}
K. Hashimoto, K. Murata, and R. Yoshii, Out-of-time-order correlators in quantum mechanics, J. High Energ. Phys. {\bf 10}, 138 (2017).

\bibitem{Li2017}
J. Li, R. H. Fan, H. Y. Wang, B. T. Ye, B. Zeng, H. Zhai, X. H. Peng, and J. F. Du, Measuring out-of-time-order correlators on a nuclear magnetic resonance quantum simulator, Phys. Rev. X {\bf 7}, 031011 (2017).

\bibitem{Mata2018}
I. Garc\'ia-Mata, M. Saraceno, R. A. Jalabert, A. J. Roncaglia, and D. A. Wisniacki, Chaos signatures in the short and long time behavior of the out-of-time ordered correlator, \prl {\bf 121}, 210601 (2018).

\bibitem{Zhao24prr}
W. L. Zhao, G. L. Li, and J. Liu, Phase modulation of directed transport, energy diffusion and quantum scrambling in a Floquet non-Hermitian system, Phys. Rev. Research {\bf 6}, 033249 (2024).


\bibitem{Chirikov91}
B. V. Chirikov, in Chaos and Quantum Mechanics, Les Houches Lecture Series Vol. 52 edited by M.-J. Giannoni, A. Voros, and J. Zinn-Justin (Elsevier Science, 1991), pp. 443-545.

\bibitem{Noronha22prb}
F. Noronha, J. A. S. Louren{\c o}, and Tommaso Macr{\` i}, Robust quantum boomerang effect in non-Hermitian systems, Phys. Rev. B {\bf 106}, 104310 (2022).

\bibitem{Tessieri21pra}
L. Tessieri, Z. Akdeniz, N. Cherroret, D. Delande, and P. Vignolo, Quantum boomerang effect: Beyond the standard Anderson model,
Phys. Rev. A {\bf 103}, 063316 (2021).

\bibitem{Sajjad22prx}
R. Sajjad, J. L. Tanlimco, H. Mas, Alec Cao, E. Nolasco-Martinez, E. Q. Simmons, F. L. N. Santos, P. Vignolo, T. Macr{\` i}, and D. M. Weld, Observation of the Quantum Boomerang Effect, Phys. Rev. X {\bf 12}, 011035 (2022).

\bibitem{Grempel84pra}
D. R. Grempel, R. E. Prange, and S. Fishman, Quantum dynamics of a nonintegrable system,
Phys. Rev. A {\bf 29}, 1639 (1984).

\bibitem{NoteSaturation}
The saturated values of the mean energy can be quantified by its time-averaged value, i.e., $\overline{\langle p^2 \rangle} = \sum_{j=1}^{N} \langle p^2(t_j) \rangle / N$, where $N$ is the total number of kicking times. In numerical simulations, we find that using one thousand kick periods $N=1000$ can ensure the good approximation of the saturated values $\overline{\langle p^2 \rangle}$.



\bibitem{JWang10pre}
J. Wang and J. B. Gong, Generating a fractal butterfly Floquet spectrum in a class of driven SU(2) systems, Phys. Rev. E {\bf 81}, 026204 (2010).


\bibitem{Suraj24pla}
K. S. Suraj, A. Kenfack, C. A. Akosa, and G. Tatara, Directed transport of Bose-Einstein Condensates with kicked interactions, Physics Letters A {\bf 497}, 129333 (2024).


\end{thebibliography}
\end{document}